\documentclass[10pt]{article}
\usepackage{latexsym}
\usepackage{amssymb}
\usepackage{amsmath}
\usepackage{amscd}
\usepackage{amsthm}
\usepackage[left=1.5cm,top=1.5cm,right=1.5cm,bottom=1.5cm]{geometry}
\usepackage{graphicx}
\usepackage{textcomp}
\usepackage{colortbl}
\usepackage{subcaption}
\usepackage[normalem]{ulem}
\usepackage{hyperref}
\usepackage[font={footnotesize,it}]{caption}
\usepackage{multirow}
\begin{document}
\begin{center}
\large{\bf{Constraining Bianchi Type I Universe With Type Ia Supernova and H(z) Data}} \\
\vspace{10mm}
\normalsize{Hassan Amirhashchi$^{1}$, Soroush Amirhashchi$^{2}$\\
\vspace{5mm}
\normalsize{$^{1}$Department of Physics, Mahshahr Branch, Islamic Azad University,  Mahshahr, Iran \\
\normalsize{$^{2}$Department of Statistics, Shahid Beheshti University, Tehran, Iran} \\	
E-mail:$^{1}$h.amirhashchi@mhriau.ac.ir},~~~$^{2}$ soroush.amirhashchi@gmail.com} \\
\end{center}
\begin{abstract}

We use recent 36 observational Hubble data (OHD) in the redshift range $0.07\leq z\leq 2.36$, latest \textgravedbl joint light curves\textacutedbl (JLA) sample, comprised of 740 type Ia supernovae (SNIa) in the redshift range $0.01\leq z \leq 1.30$, and their joint combination datasets to constrain anisotropic Bianchi type I (BI) dark energy (DE) model. To estimate model parameters, we apply Hamiltonian Monte Carlo technique. We also compute the covariance matrix for BI dark energy model by considering different datasets to compare the correlation between model parameters. To check the acceptability of our fittings, all results are compared with those obtained from 9 year WMAP as well as Planck (2015) collaboration. Our estimations show that at 68\% confidence level the dark energy equation of state (EOS) parameter for OHD or JLA datasets alone varies between quintessence and phantom regions whereas for OHD+JLA dataset this parameter only varies in phantom region. It is also found that the current cosmic anisotropy is of order $\sim10^{-3}$ which imply that the OHD and JLA datasets do not put tight constraint on this parameter. Therefore, to constraint anisotropy parameter, it is necessary to use high redshif dataset namely cosmic microwave background (CMB). Moreover, from the calculation of $p$-value associated with $\chi^{2}$ statistic we observed that non of the  $\omega \mbox{BI}$ and flat $\omega\mbox{CDM}$ models rule out by OHD or JLA datasets. The deceleration parameter is obtained as $q=-0.46^{+0.89 +0.36}_{-0.41 -0.37}$, $q=-0.619^{+0.12 +0.20}_{-0.095 -0.24}$, and $q=-0.52^{+0.080 +0.014}_{-0.046 -0.15}$ for OHD, SNIa, and OHD+SNIa data respectively.
\end{abstract}
%%%%%%%%%%%%%%%%%%%%%%%%%%%%%%%%%%%%%%%%%%%%%%%%%%%%%%%%%%%%%%%%%%%%%%%%%%%%%%%%
\section{Introduction}
\label{sec:intro}
The cornerstone of recent day cosmology believes that the place where we are living has no privileged position in the universe. This simple and powerful idea is called cosmological principle (CP). Mathematically one can say that there are necessarily translational symmetries from any point of space to any other which implies that space should be homogeneous (universe looks same at any point). Moreover, at enough large scales since the universe looks same in any direction, there should be rotational symmetries which imposes an isotropic property to the geometry of space. A maximally symmetric space-time satisfying the cosmological principle is given by Friedmann-Robertson-Walker (FRW) metric. From observational point of view, it is widely believed that our universe could be accurately described by FRW model as the cosmic microwave background (CBM) temperature is highly isotropic about our position. Nevertheless, the high symmetry of FRW models represents a very high degree of fine tuning of initial conditions which implies that this models are infinitely improbable in the space of all possible cosmologies. Although, the observed universe could be described by FRW models at current epoch. There are some important questions i) does our universe necessarily posses the same symmetries outside the particle event horizon? ii) Are there possible models that will fit the observations rather than FRW models? Furthermore, the recent observations indicate small variations between the intensities of cosmic microwave background (CMB) coming from different directions which may be related to the origin of structure in the universe. Of course, a more general and realistic metric posses both inhomogeneity and anisotropy properties, but in this case the exact solution of Einstein's field equations are almost impossible. Therefore, we usually simplify this general metric in following two sub classes: i) isotropic and inhomogeneous models given by  Lema\^{i}tre-Tolman-Bondi (LTB) metric \cite{ref1,ref2,ref3,ref4} ii) anisotropic and homogeneous models given by Bianchi metrics \cite{ref5}. In fact, at least, these models provide an arena for testing the accuracy of FRW models in describing our universe at the present epoch. It is worth to mention that some Bianchi models isotropize due to inflation \cite{ref6}.\\

According to observations the expansion rate of universe is accelerating at present epoch \cite{ref7,ref8,ref9}. In the context of General Relativity (GR), there must be an extra component in the cosmic fluid which acts against gravity. In literature, this extraordinary component is called dark energy (DE). Since we still could not detect any interaction between DE and ordinary matter, in spite of many efforts, our informations about this component is quite less. Fortunately, the nature of DE could be investigated through its equation of state parameter (EOS) which defined as the ratio of pressure to energy density $\omega=p/\rho$. Recent 9 year WMAP \cite{ref10} and Planck (2015) \cite{ref11} collaboration results, at $68\%$ confidence level (CL), show that the EOS parameter varies in the ranges $-1.162<\omega_{X}<-0.983$ and $-1.099<\omega_{X}<-0.944$ respectively ($\omega_{X}$ refers to the dark energy EOS parameter). The dark energy EOS parameter could be considered as a constant parameter (i.e $\omega_{X}=-1$) described by cosmological constant or a dynamical time varying function of time (or equivalently redshift) which could be described by scalar fields.  As interval $-1/3>\omega_{X}>-1$ is called quintessence region, phantom region is indicated by $\omega_{X} < -1$. While cosmological constant scenario faces two serious theoretical problems namely the fine-tuning and coincidence problem \cite{ref12, ref13}, phantom scenario suffers from ultraviolet quantum instabilities \cite{ref14} and quintessence does not match with recent observations \cite{ref7,ref8,ref9} which indicate the possibility of crossing phantom divide line (PDL) at $68\%$ CL. The thing which we almost know precisely is that at matter dominated era the cosmic expansion was decelerating but at a certain redshift called transition redshift, the dark energy dominated over the universe and hence the expansion of the universe has been changed from decelerating to accelerating phase. We can investigate this phase transition by tracing the sign change of the universal deceleration parameter $q(z)$ in the history of cosmic evolution. In general, basic characteristics of the cosmological evolution could be expressed in terms of the Hubble parameter $H_{0}$ and the deceleration parameter $q_{0}$ \cite{ref15}.\\

In the recent past, Zibin \cite{ref16,ref17}, Valkenburg et al \cite{ref18}, Zumalacarregui et al \cite{ref19} and Tokutake et al \cite{ref20} have investigated dark energy models under the framework of inhomogeneous LTB spacetime in different physical contexts. Recently the authors of Ref \cite{ref21} have studied some DE models in the scope of LTB spacetime. Some important applications of DE models in the scope of anisotropic Bianchi spacetimes are given in Refs. \cite{ref22,ref23,ref24,ref25,ref26,ref26,ref27,ref28} (for review, see \cite{ref29,ref30}). In our previous work, we have studied viscous dark energy in the scope of Bianchi type V spacetime \cite{ref31}. It is worth noting that for more than five decades there have been considerable studies of CMB temperature in spatially homogeneous universes which have used the observed temperature anisotropy to place constraint on the overall anisotropy of the cosmic expansion \cite{ref32,ref33}. Motivated by above researches, in this paper, we confine ourselves to investigate dark energy in the scope of Bianchi type I ($\omega$BI) universe through the recent 36 observational Hubble data (OHD) in the intermediate redshift $0.07\leq z\leq 2.36$ compiled by Yu et al \cite{ref34}, \textgravedbl latest joint light curves\textacutedbl (JLA) dataset \footnote{All data used are available on	http://supernovae.in2p3.fr/sdss-snls-jla/ReadMe.html} comprised of 740 type Ia supernovae in the redshift range $z \epsilon [0.01, 1.30]$ and their joint combination. Note that JLA dataset provides model-independent apparent magnitudes instead of model-dependent distance moduli whereas several SN datasets such as Union provide cosmological distance moduli that are derived assuming a flat $\Lambda$CDM model and hence can not be applied to other models such as Bianchi spacetimes. We estimate all parameters of BI dark energy model (some other parameters are also derived from fit parameters) by the aide of Hamiltonian Monte Carlo (HMC) technique and compare our results to the 9years WMAP \& Planck(2015) to evaluate the robustness of our fits. The paper is structured as follows: In section~\ref{sec:theor} we briefly discuses the theoretical DE models. Section~\ref{sec:method} deals with the summary of computational technique we have used to fit parameters to data. In section~\ref{sec:res} we study $\omega$BI dark energy model and fit its parameters to OHD, JLA, and their joint combination datasetes. In section~\ref{sec: God} we used the Akaike information criterion (AIC) and Bayes factor ($\Psi$) to better compare the considered model. Finally, we summarize our findings and conclusions in section~\ref{sec: con}.
%%%%%%%%%%%%%%%%%%%%%%%%%%%%%%%%%%%%%%%%%%%%%%%%%%%%%%%%%%%%%%%%%%%%%%%%%%%%%%%%%%%%%%
%%%%%%%%%%%%%%%%%%%%%%%%%%%%%%%%%%%%%%%%%%%%%%%%%%%%%%%%%%%%%%%%%%%%%%%%%%
%%%%%%%%%%%%%%%%%%%%%%%%%%%%%%%  SECTION 2  %%%%%%%%%%%%%%%%%%%%%%%%%%
\section{Theoretical Models}
\label{sec:theor}
In synchronous coordinate system, we construct the following general $(N + 1)$-dimensional {\bf inhomogeneous} and {\bf anisotropic} Lorentzian spacetime with the metric
\begin{equation}
\label{eq1}
ds^{2} = -dt^{2} + \delta_{ij}g_{ij}dx^{i}dx^{j}, ~~~~i,j=1,2,\dots N,
\end{equation}
where $g_{ij}$ are functions of $(t,x^{1},x^{2},x^{3})$ and $t$ refers to the cosmological (or cosmic) time. In 4-dimensions, we could generate FRW and Bianchi type I models from above equation as
\begin{equation}
\label{eq2} \begin{cases}
\mbox{if}~~\delta_{ij}g_{ij}=a^{2}(t,x)~~ &\mbox{FRW model} \\
\mbox{if}~~\delta_{ij}g_{ij}=a^{2}_{ij}(t,x)~~  &\mbox{BI model}\\
\end{cases}
\end{equation}
Above relations show that for FRW universe all three metric potentials are equal (i.e $g_{11}=g_{22}=g_{33}=a^{2}(t,x)$) which demonstrates an {\bf isotropic} but {\bf inhomogeneous} spacetime whereas for BI model, metric components are different functions ($g_{11}=A^{2}(t,x), g_{22}=B^{2}(t,x),g_{33}=C^{2}(t,x)$) which indicates an {\bf anisotropic} and {\bf inhomogeneous} spacetime. It is worth to note that in BI case, the average scale factor is defined as $a=(ABC)^{1/3}$. In an inhomogeneous universe, metric components are function of time and spatial coordinates. But, for simplicity, we assume that in both FRW and BI models, metric components are functions of time only. Hence, FRW describes a homogeneous and isotropic universe which obeys the cosmological principle (CP) whereas BI is homogeneous but anisotropic which violates CP. We consider the possible constituents  of the universe to be in the perfect fluids form, meaning that we neglect the effect of viscosity or heat flow. Under this condition, the perfect fluid energy-momentum tensor could be written as \\
\begin{equation}
\label{eq3}
T_{ij}=\mbox{diag}(-\rho c^{2},p,p,p),
\end{equation}
where $\rho$ is the total energy density, $p$ is pressure and $c$ is the speed of light. The Einstein's field equations ( in gravitational units $8\pi G = c = 1 $) read as
\begin{equation}
\label{eq4} R_{ij} - \frac{1}{2} R g_{ij} = T_{ij}.
\end{equation}
For general metric (\ref{eq1}), the $0-0$ and the $i-i$ components of Einstein's equation lead to the following equations \cite{ref35}
\begin{equation}
\label{eq5}
\left(\frac{\dot{a}}{a}\right)^{2}=\frac{1}{3}(\rho_{m}+\rho_{X}+\rho_{r})+\frac{\tilde{k}}{a^{n}},
\end{equation}
and
\begin{equation}
\label{eq6} \frac{\ddot{a}}{a}=-\frac{1}{6}(\rho_{m}+\rho_{X}+\rho_{r}+3p).
\end{equation}
respectively. Here $(\rho_{m}, \rho_{r}, \rho_{X})$ are the DE, DM, and radiation energy densities and $\tilde{k}=(k, A_{0})$ for FRW and BI models respectively, where $k$ stands for curvature and $A_{0}$ indicates the anisotropy amount of BI model \footnote{In Ref \cite{ref35} it has been shown that the anisotropy parameter in BI model decays as $A=A_{0}a^{-6}$.}. The density fractions $ \Omega_{m}, \Omega_{r}, \Omega_{X}, \Omega_{k}$, and $\Omega_{A}$ are defined by

\begin{equation}
\label{eq7} \Omega_{m}=\frac{\rho_{m}}{3H^{2}},~~~~ \Omega_{r}=\frac{\rho_{r}}{3H^{2}},~~~~ \Omega_{X}=\frac{\rho_{X}}{3H^{2}},~~~~\Omega_{k}=-\frac{k}{H^{2}},~~~~\Omega_{A}=-\frac{A_{0}}{H^{2}}.
\end{equation}
Therefore, from (\ref{eq5}) the Hubble parameter H(z) is
\begin{equation}
\label{eq8} H(z)^{2}=H_{0}^{2}\left[\Omega_{m}(1+z)^{3}+\Omega_{r}(1+z)^{4}+\Omega_{X}(1+z)^{3(1+\omega_{X})}+\Omega_{\tilde{k}}(1+z)^{n}\right],
\end{equation}
where $\omega_{X}=p_{X}/\rho_{X}$ is the equation of state parameter of DE fluid (note that, as usual, $p_{m}=0$ which imply $\omega_{m}=0$) and $a=(1+z)^{-1}$. Requiring the consistency of (\ref{eq8}) at $z=0$, gives
\begin{equation}
\label{eq9} \begin{cases}
\Omega_{m}+\Omega_{r}+\Omega_{X}+\Omega_{k}=1~~ &\mbox{FRW model} \\
\Omega_{m}+\Omega_{r}+\Omega_{X}+\Omega_{A}=1~~  &\mbox{BI model}\\
\end{cases}
\end{equation}
From  eq (\ref{eq8})the possible cosmologies that could be considered in our study are shown in Table~\ref{tab:1}.\\
%%%%%%%%%%%%%%%%%%%%%%%%%%%%%%%%%%%%%%%%%%%%Table 1%%%%%%%%%%%%%%%%%%%%%%%%%%%%%%%%%%%%%%%%%%%%%%%%%%%%%%%%%%%
%%%%%%%%%%%%%%%%%%%%%%%%%%%%%%%%%%%%%%%%%%%%%%%%%%%%%%%%%%%%%%%%%%%%%%%%%%%%%%%%%%%%%%%%%%%%%%%%%%%%%%
\begin{table}[ht]
\caption{Three possible cosmological models which could be derived from eq (\ref{eq8})}
\centering
\begin{tabular} {|>{\columncolor[gray]{0.8}}cccc|}
\hline
Parameter    & Flat $\omega$CDM & Non-Flat $\omega$CDM & $\omega$BI  \\[0.5ex]
\hline
\hline
$n$ &  =2 &   =2 &  =6\\ 
		
$H_{0}$ &  \checkmark   & \checkmark & \checkmark\\
		
$\Omega_{m}$ &  \checkmark   & \checkmark & \checkmark\\
		
$\Omega_{X}$ & \checkmark  &  \checkmark & \checkmark \\ 
		
$\Omega_{r}$ & \checkmark & \checkmark & \checkmark \\
		
$\Omega_{\tilde{k}}$ & \texttimes & $=\Omega_{k}$ & $=\Omega_{A}$ \\
		
$\omega_{X}$ &  \checkmark   & \checkmark & \checkmark\\[0.5ex]
		
\hline
\end{tabular}
\label{tab:1}
\end{table}
%%%%%%%%%%%%%%%%%%%%%%%%%%%%%%%%%%%%%%%%%%%%%%%%%%%%%%%%%%%%%%%%%%%%%%%%%%%%%%%%%%%%%%%%%%%%%%%%%
%%%%%%%%%%%%%%%%%%%%%%%%%%%%%%%%%%%%%%%%%%%%%%%%%%%%%%%%%%%%%%%%%%%%%%%%%%%%%%%%%%%%%%%%%%%%%%%

In the history of cosmic evolution, initially, our universe was undergoing a decelerating expansion, then
at a certain redshift (time) $z_{t}$ dark energy dominates over universe which in turn changes the expansion phase from decelerating to accelerating. The transition redshift is obtained by condition $q(z_{t})=\ddot{a_{t}}=0$. The deceleration parameter is defined as
\begin{equation}
\label{eq10} q(z)=-\frac{1}{H^{2}}\left(\frac{\ddot{a}}{a}\right)=\frac{(1+z)}{H(z)}\frac{dH(z)}{dz}-1,
\end{equation}
which in turn gives 
\begin{equation}
\label{eq11} q(z)=\frac{1}{2}\bigg[2-\Omega_{m}(1+z)^{3}-(1-3\omega_{X})\Omega_{X}(1+z)^{3(1+\omega_{X})}-n\Omega_{\tilde{k}}(1+z)^{n}\bigg],
\end{equation}
Therefore, the transition redshif could be obtained by numerical solution of the following equation
\begin{equation}
\label{eq12} \Omega_{m}(1+z)^{3}+(1-3\omega_{X})\Omega_{X}(1+z)^{3(1+\omega_{X})}+n\Omega_{\tilde{k}}(1+z)^{n}=2.
\end{equation}
For all three models, one also could obtain the age of the Universe in terms of the redshift $z$ as
\begin{equation}
\label{eq13} dt=\frac{da}{aH}\Rightarrow t_{0}=\int_{0}^{\infty}\dfrac{dz}{(1+z)H},
\end{equation}
which in turn gives (by the aid of eq (\ref{eq8}))
\begin{equation}
\label{eq14} t_{0}=H_{0}^{-1}\int_{0}^{\infty}\dfrac{dz}{(1+z)\sqrt{\Omega_{m}(1+z)^{3}+\Omega_{r}(1+z)^{4}+\Omega_{X}(1+z)^{3(1+\omega_{X})}+\Omega_{\tilde{k}}(1+z)^{n}}}.
\end{equation}
In the next section, we fit $\omega$BI dark energy model to the OHD, JLA, and their joint combination. All estimated results will be compared to those obtained by other researchers as will as 9 years WMAP \cite{ref10} and Planck 2015 collaboration \cite{ref11} as shown in Table~\ref{tab:2}. Since recent observations show that the radiation density, $\Omega_{r}$, is of order $\sim10^{-5}$, we shall omit this parameter from our analysis.
%%%%%%%%%%%%%%%%%%%%%%%%%%%%%%%%%%%%%%%%%%%%Table 2%%%%%%%%%%%%%%%%%%%%%%%%%%%%%%%%%%%%%%%%%%%%%%%%%%%%%%%%%%%
%%%%%%%%%%%%%%%%%%%%%%%%%%%%%%%%%%%%%%%%%%%%%%%%%%%%%%%%%%%%%%%%%%%%%%%%%%%%%%%%%%%%%%%%%%%%%%%%%%%%%%
\begin{table}[ht]
\caption{Results from 9years WMAP and Planck 2015 collaboration for $\Lambda$CDM model at 1$\sigma$ confidence level.}
\centering
\setlength{\tabcolsep}{25pt}
\scalebox{0.8}{
\begin{tabular} {ccc}
\hline
Parameter    & WMAP+eCMB+BAO+H$_{0}$& TT+TE+EE+lensing+BAO+JLA+H$_{0}$ \\[0.5ex]
\hline
\hline{\smallskip}
$H_{0}$ &  $68.92^{+0.94}_{-0.95}$ &    $67.74\pm0.46$ \\[0.2cm] 
			
$\Omega_{m}$ &  $0.2855^{+0.0096}_{-0.0097}$ &  $0.3089\pm0.0062$ \\[0.2cm] 
			
$\Omega_{X}$ & $0.717\pm0.011$ & $0.6911\pm0.0062$\\[0.2cm] 
			
$\Omega_{k}$ & $-0.0027^{+0.0039}_{-0.0038}$ &$0.0008^{+0.0040}_{-0.0039}$\\[0.2cm] 
			
$\omega_{X}$ & $-1.073^{+0.090}_{-0.089}$ & $-1.019^{+0.075}_{-0.080}$\\[0.2cm] 			
			
$t_{0}$ & $13.88\pm0.16$& $13.799\pm0.021$ \\[0.5ex]
			
\hline\hline
\end{tabular}}
\label{tab:2}
\end{table}

%%%%%%%%%%%%%%%%%%%%%%%%%%%%%%%%%%%%%%%%%%%%%%%%%%%%%%%%%%%%%%%%%%%%%%
%%%%%%%%%%%%%%%%%%%%%%%%%%%%%%%   SECTION 3  %%%%%%%%%%%%%%%%%%%%%%%%%
\section{Data Sets And Method}
\label{sec:method}
In what follows we use {\bf NUTS} sampler which is an extension of {\bf Hamiltonian Monte Carlo (HMC)} algorithm to generate MCMC chains and place constraints on cosmological parameters of $\omega$BI dark energy model. To do so, we use independent observables that are:
\begin{itemize}
\item Observational Hubble dataset (OHD): including $36H(z)$ datapoints (see Table~\ref{tab:3}) in the redshift range $0.07\leq z\leq 2.36$ \cite{ref34} (note that because of the partial overlap of the WiggleZ and BOSS spatial regions (see Beutler et al \cite{ref36}), we drop the three Blake et al \cite{ref37} WiggleZ radial BAO points from Table 1 of Farooq et al \cite{ref38} but include the recent redshift $z = 0.47$ cosmic chronometric measurement \cite{ref39}). Generally, the OHD data is categorized into the following two categories i) BAO based data in which we usually model the redshift space distortions and assume an acoustic scale in a specific model and ii) cosmic chronometric (CC) based data in which we use the most massive and
passively evolving galaxies based on the galaxy differential age method. As mentioned by Yu et al \cite{ref34}, 31 data of this $36H(z)$ measurements are determined using the CC technique, 3 data from the radial BAO signal in the galaxy distribution and 2 data from the BAO signal in the Lyman α forest distribution alone or cross correlated with QSOs. For this dataset, we minimize the following log marginal likelihood function
\begin{equation}
\label{eq15} \ln\mathcal{L}=-\frac{1}{2}\sum_{i,j}^{N}\left[H_{th}(z)-H_{obs}(z_{i})\right]({\bf C}^{-1})_{ij}\left[H_{th}(z)-H_{obs}(z_{j})\right],
\end{equation}
where ${\bf C}^{-1}_{ij}$ is the inverse of covariance matrix of the observed data. It should be noted that three galaxy distribution radial BAO $H(z)$ measurements are correlated and their correlation matrix is given by (see Ref \cite{ref48} for details)
\begin{equation}
\label{eq16} {\bf c_{1}}=
\begin{pmatrix}
3.65 & 1.78 & 0.93  \\
1.78 & 3.65 & 2.20\\
0.93 & 2.20 & 4.45 
\end{pmatrix}.
\end{equation}
Also the covariance matrix for other uncorrelated data is simply given by ${\bf c_{2}}=\mbox{diag}(\sigma_{H}^{2})_{i}$ (see Table~\ref{tab:3}). Hence the total covariance matrix, in eq \ref{eq15}, is 

\begin{equation}
{\bf C}=
\begin{bmatrix}
{\bf c_{1}} & {\bf 0} \\
{\bf 0} & {\bf c_{2}}
\end{bmatrix}.
\end{equation}

\item Supernovae Type Ia dataset (JLA): comprised of $740$ type Ia supernovae in the redshift range $0.01\leq z\leq 1.30$ \cite{ref49}. In case of SNIa dataset, we minimize

\begin{equation}
\label{eq17} \ln\mathcal{L}=-\frac{1}{2}\sum_{i,j}^{N}\left[\mu_{th}(z)-\mu_{obs}(z_{i})\right]({\bf C}^{-1})_{ij}\left[\mu_{th}(z)-\mu_{obs}(z_{j})\right],
\end{equation}
where the covariance matrix $ {\bf C} $ is also available in \cite{ref49} and the predicted distance modulus, $\mu(z)$, for a flat space-time may be given by
\begin{equation}
\label{eq18} \mu(z)=5\log_{10}\left[3000 (1+z) \int_{0}^{\infty}\frac{dz}{E(z)}\right]+25-5\log_{10}(h).
\end{equation}
Here $E(z)=H(z)/H_{0}$ is the reduced Hubble parameter given by eq (\ref{eq8}) and $H_{0}=100h~ km~ s^{-1} Mpc^{-1}$. It is clear that $h$ is an additive constant and hence marginalizing over it does not affect the SNe results. 
%It is worth to mention that for this dataset the covariance matrix is given by ${\bf C}_{ij}=\mbox{diag}(\sigma_{i}^{2})$, where $\sigma_{i}^{2}$ is the variance of any single data.
\end{itemize} 
Since these two datasets are assumed to be independent, the total likelihood could be defined as the product of the likelihoods of the single datasets. The total likelihood is given by
\begin{equation}
\label{eq19} \mathcal{L}_{tot}= \mathcal{L}_{OHD}\times\mathcal{L}_{SNIa}.
\end{equation}
Note that joint combination could increase the sensitivity of our estimates. In following section, we estimate parameters of flat $\omega$BI dark energy model. We also derive (estimate) transition redshift $z_{t}$, deceleration parameter $q$, and age of the universe $t_{0}$ for this model. The prior for all parameters of model are assumed to be uniform.
%%%%%%%%%%%%%%%%%%%%%%%%%%%%%%%%%%%%%%%%%%%%%%%%%%Table 3%%%%%%%%%%%%%%%%%%%%%%%%%%%%%%%%%%%%%%%%%%%%%%%%%%%%%
%%%%%%%%%%%%%%%%%%%%%%%%%%%%%%%%%%%%%%%%%%%%%%%%%%%%%%%%%%%%%%%%%%%%%%%%%%%%%%%%%%%%%%%%%%%%%%%%%%%%%%%%%%%%%%
\begin{table}[ht]
\caption{Hubble parameter versus redshift data.}
\centering
\setlength{\tabcolsep}{25pt}
\scalebox{0.7}{
\begin{tabular} {cccc}
\hline
\hline
$H(z)$    &  $\sigma_{H}$   &  $z$  & Reference\\[0.5ex]
			
\hline{\smallskip}
69 &      19.6     & 0.070   & \cite{ref47}\\
			
69 &      12       & 0.090   & \cite{ref43} \\
			
68.6 &    26.2     & 0.120   & \cite{ref47}\\
			
83 &      8        & 0.170   & \cite{ref40} \\
			
75 &      4        & 0.179   & \cite{ref42} \\
			
75 &      5        & 0.199   & \cite{ref42} \\
			
72.9 &    29.6     & 0.200   & \cite{ref43} \\
			
77 &      14       & 0.270   & \cite{ref40} \\
			
88.8 &    36.6     & 0.280   & \cite{ref43} \\ 
			
83  &     14       &0.352    & \cite{ref42} \\
			
81.9  &   1.9      &0.380    & \cite{ref48} \\
			
83  &     13.5     &0.3802   & \cite{ref47} \\
			
95  &     17       &0.400    & \cite{ref40} \\
			
77  &     10.2     &0.4004   & \cite{ref47} \\
			
87.1  &   11.2     &0.4247   & \cite{ref47} \\
			
92.8  &   12.9     &0.4497   & \cite{ref47} \\
			
89    &   50       &0.47     & \cite{ref39} \\
			
80.9  &   9        &0.4783    & \cite{ref47} \\
			
97 &      62       & 0.480   &\cite{ref41}  \\
			
90.8  &   1.9       &0.510    & \cite{ref48} \\
			
104 &     13       &0.593    & \cite{ref42} \\
			
97.8  &   2.1      &0.610    & \cite{ref48} \\
			
92 &      8        &0.680    &\cite{ref42}  \\
			
105 &     12       &0.781    & \cite{ref42} \\
			
125 &     17       &0.875    & \cite{ref42} \\
			
90 &      40       &0.880    & \cite{ref41} \\
			
117 &     23       &0.900    & \cite{ref40} \\ 
			
154 &     20       &1.037    & \cite{ref42} \\
			
168 &     17       & 1.300   & \cite{ref40} \\ 
			
160  &    33.6     &1.363    & \cite{ref46} \\
			
177 &     18       &1.430    & \cite{ref40} \\
			
140 &     14       &1.530    & \cite{ref40} \\
			
202 &     40       &1.750    & \cite{ref40} \\
			
186.5 &   50.4     &1.965    & \cite{ref46} \\
			
222  &    7        &2.340    & \cite{ref45} \\ 
			
227  &    8        &2.360    & \cite{ref44} \\ 
			
\hline
\hline
\end{tabular}}
\label{tab:3}
\end{table}
%%%%%%%%%%%%%%%%%%%%%%%%%%%%%%%%%%%%%%%%%%%%%%%%%%%%%%%%%%%%%%%%%%%%%%%%%%%%%%%%%%%%%%%%%%%%%%%
%%%%%%%%%%%%%%%%%%%%%%%%%%%%%%%%%%%%%%%%%%%%%%%%%%%%%%%%%%%%%%%%%%%%%%%%%%%%%%%%%%%%%%%%%%%%%%%%%%
\section{Results and discussion}
\label{sec:res}
Dark energy $\omega$BI model has five unknown parameters to be estimated from $36H(z)$, JLA, and their joint combination. The base parameters set for this model is
%%%%%%%%%%%%%%%%%%%%%%%%%%%%%%%%%%%%%%%%%%%%%%%%%%%%%% Figure 1 %%%%%%%%%%%%%%%%%%%%%%%%%%%
%%%%%%%%%%%%%%%%%%%%%%%%%%%%%%%%%%%%%%%%%%%%%%%%%%%%%%%%%%%%%%%%%%%%%%%%%%%%%%%%%%%%%%%%%%
\begin{figure}[h!]
\centering
\includegraphics[width=18cm,height=18cm,angle=0]{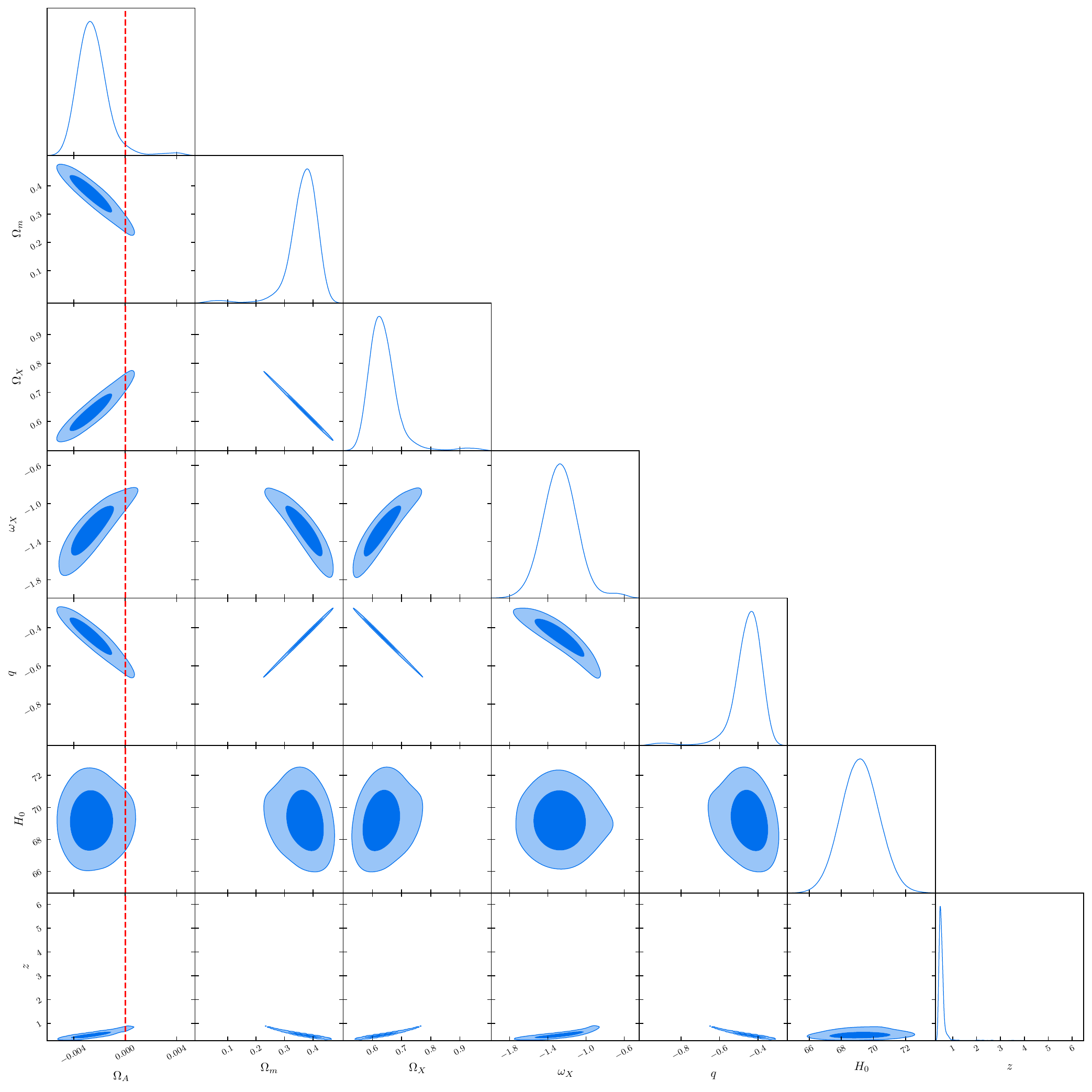}
\caption{\label{fig1} One-dimensional marginalized distribution and two-dimensional contours with $68\%$ CL and $95\%$ CL for $\omega$BI model Using {\bf OHD+JLA} dataset.}

\end{figure}
%%%%%%%%%%%%%%%%%%%%%%%%%%%%%%%%%%%%%%%%%%%%%%%%%%%%%%%%%%%%%%%%%%%%%%%%%%%%%%%%%%%%%%%%%
%%%%%%%%%%%%%%%%%%%%%%%%%%%%%%%%%%%%%%%%%%%%%%%%%%%%%%%%%%%%%%%%%%%%%%%%%%%%%%%%%%%%%%%%%
\begin{equation}
\label{eq23} {\bf \theta}=\{\Omega_{m},\Omega_{X}, \Omega_{A},\omega_{X}, H_{0}\}.
\end{equation}
Table~\ref{tab:4} demonstrates the results of our statistical analysis for $\omega$BI dark energy model using OHD, JLA, and their joint combination. Also, we have depicted the two-dimensional joint contours (OHD+JLA), at $68\%$ and $95\%$ CL, and the one-dimensional marginalized posterior distributions for cosmological parameters in Figure.~\ref{fig1}. From Table~\ref{tab:4} we observe that, at $68\%$ confidence level, when we constraint model over OHD or JLA datasets alone, the dark energy EOS parameter varies between quintessence and phantom regions (for OHD $-2.13\leq \omega_{X}\leq -0.81$ and for JLA $-1.36\leq \omega_{X}\leq -0.84$) whereas for OHD+JLA dataset this parameter only varies in phantom region ($-1.46\leq \omega_{X}\leq -1.08$). Therefore, it is obvious that our results support phantom dark energy scenario in BI dark energy model. This result is in agreement with 9 years WMAP \cite{ref10}\& Planck (2015) collaboration \cite{ref11}. From Table~\ref{tab:4} we observe that there is about $\sim0.8\sigma$ of a deviation of the EOS parameter from the cosmological constant from OHD data (this is only $\sim 0.5\sigma$ from JLA). This deviation from $\omega_{\Lambda}=-1$ could be partly due to systematics in these data.\\
Our joint analysis indicates that the anisotropy parameter vary in interval $-38\times10^{-4}\leq\Omega_{A}\leq-16\times10^{-4}$ at $68\%$ error which is 100 times larger than the level  of anisotropies, $\sim 10^{-5}$, observed in the CMB measurements. This result shows that using these two datasets is not enough to constrain anisotropy parameter, $\Omega_{A}$, in BI universe. It is worth mentioning that $H(z)$ data is not sensitive to the behavior of cosmological spatial inhomogeneities \cite{ref50}. Hence, more precise measurements of $H(z)$ at higher redshifts are needed in order to put tighter constraints on the anisotropy parameter $\Omega_{A}$. Also we observe that for the joint combination, OHD+JLA, whilst the estimated value of the current expansion rate of the universe i.e $H_{0}$ is in good agreement to those obtained by Adel et al, Planck, ($67.74\pm0.46$)\cite{ref11}, Hinshaw et al, WMAP, ($68.92\pm0.84$) \cite{ref10} and specially with that obtained by Chen \& Ratra ($68\pm2.8$) \cite{ref52}, but deviates from what reported by Riess et al ($73.4\pm1.74$)\cite{ref51}. Figures.~\ref{fig2},~\ref{fig3} depict the robustness of our fits. From Figure.~\ref{fig2} we observe that the joint dataset gives raise to a quite better fit. It is worth noting that although JLA dataset by itself is not sensitive to the expansion rate $H_{0}$, however, in the joint analysis it constrains other parameters of the model which in turn affect the estimate of $H_{0}$. This is why in Table~\ref{tab:4} we observe a change in the value of $H_{0}$ when fitting model to the joint OHD+JLA dataset. From the $H(z)$ data we find evidence for the cosmological deceleration-acceleration transition to have taken place at a redshift $z_{t}=0.72\pm0.14$. This computed $z_{t}$ is in good agreement, at $68\%$ CL, with those obtained by Farooq et al \cite{ref38} determination of $z_{t}=0.72\pm0.05$, Busca et al \cite{ref53} determination of $z_{t}=0.82\pm0.08$ and those reported in Tables~\ref{tab:1} and \ref{tab:2} of Ref \cite{ref54}. Moreover, from The join combination of $H(z)$ and SNIa datasets we find that the change of the BI expansion phase from deceleration to acceleration takes place at redshift $z_{t}=0.57^{+0.0037}_{-0.15}$ which is in good agreement with the results obtained by Vargas dos Santos et al \cite{ref55}.\\
%%%%%%%%%%%%%%%%%%%%%%%%%%%%%%%%%%%%%%%%%%%%Table 3%%%%%%%%%%%%%%%%%%%%%%%%%%%%%%%%%%%%%%%%%%%%%%%%%%%%%%%%%%%
%%%%%%%%%%%%%%%%%%%%%%%%%%%%%%%%%%%%%%%%%%%%%%%%%%%%%%%%%%%%%%%%%%%%%%%%%%%%%%%%%%%%%%%%%%%%%%%%%%%%%%
\begin{table}[h!]
\caption{Results from the fits of the flat $\omega$BI model to the data at $68\%$ confidence level.}
\centering
\setlength{\tabcolsep}{25pt}
\scalebox{0.8}{
\begin{tabular}{c p{1cm} cccc}
\hline
\hline
& & Parameter & OHD & JLA & OHD+JLA  \\
\hline
&\multirow{3}{*}{Fit parameters} &
$H_{0}$ &  $69.49\pm 0.70$ & - &  $69.2\pm 1.2$\\[.2cm]
& &$\Omega_{m}$ &  $0.370^{+0.098}_{-0.40}$   & $0.315^{+0.16}_{-0.084}$ &  $0.364^{+0.054}_{-0.031}$\\[.2cm]
& &$\Omega_{A}$ & $-0.00275^{+0.00095}_{-0.0019}$  &  $-0.0037^{+0.0098}_{-0.016}$  & $-0.00246^{+0.00086}_{-0.0013}$ \\[.2cm]
& &$\omega_{X}$ & $-1.470\pm 0.66$ & $-1.12^{+0.28}_{-0.24}$ & $-1.27\pm 0.19$\\[.2cm]
\cline{2-6}
& \multirow{3}{*}{Derived parameters} &
$\Omega_{X}$ & $0.633^{+0.039}_{-0.096}$ & $0.688^{+0.075}_{-0.15}$ & $0.638^{+0.030}_{-0.053}$ \\[.2cm]
&&$z_{t}$ & $0.4097\pm 0.0089$  & $0.80^{+0.070}_{-0.61}$  & $0.57^{+0.0073}_{-0.15}$ \\[.2cm] 
& &$t_{0}$ & $11.5\pm0.72$ & $12.23^{+0.34}_{-0.22}$ & $12.36^{+0.27}_{-0.12}$ \\[.2cm]
& &$q$ & $-0.359\pm 0.011$ & $-0.619^{+0.12}_{-0.095}$ & $-0.52^{+0.080}_{-0.046}$\\[.2cm]
\hline
\hline
\end{tabular}}
\label{tab:4}
\end{table}
In Figure panel~\ref{fig4} we have plotted the dependence of deceleration parameter $q(z)$ as a function of redshift $z$ for OHD (fig.~\ref{fig4a}), JLA (fig.~\ref{fig4b}), and OHD+JLA (fig.~\ref{fig4c}) at $68\%$ \& $95\%$ CL. The solid lines show the mean value of $q(z)$ and filled circles indicate the best fit values of the deceleration parameter at $z_{t}$. Our statistical analysis show that $q=-0.46^{+0.89 +0.36}_{-0.41 -0.37}$, $q=-0.619^{+0.12 +0.20}_{-0.0.095 -0.24}$, and $q=-0.52^{+0.080 +0.014}_{-0.046 -0.15}$ for H(z), SNIa, and H(z)+SNIa datasets respectively. This results are in good agreement with those reported in Refs \cite{ref35},\cite{ref53}, and \cite{ref54}. It should be pointed out that the confidence limits in Table~\ref{tab:4} are $ 95\% $ two-tail limits which are calculated using posterior samples (chains). To be more clear, for the determination of these confidence limits, we have calculated the fraction(in our case, $ 2.5\% $) of the posterior samples in each tail.\\

%%%%%%%%%%%%%%%%%%%%%%%%%%%%%%%%%%%%%%%%%%%%%%%%%%%%%% Figure 2 and 3 %%%%%%%%%%%%%%%%%%%%%%%%%%%
%%%%%%%%%%%%%%%%%%%%%%%%%%%%%%%%%%%%%%%%%%%%%%%%%%%%%%%%%%%%%%%%%%%%%%%%%%%%%%%%%%%%%%%%%%%%%%%%
\begin{figure}[h!]
\begin{minipage}[b]{0.5\linewidth}
\centering
\includegraphics[width=8cm,height=6cm,angle=0]{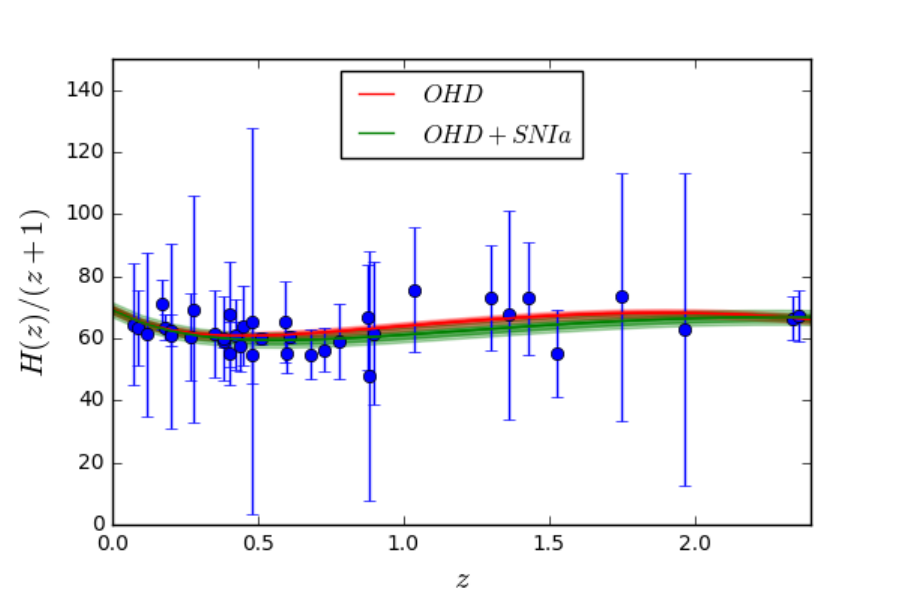} \\
\caption{\label{fig2} The Hubble rate of the flat $\omega$BI model versus the redshift $z$ at $68\%$ and $95\%$ confidence level. The points with bars indicate the experimental data summarized in Table~\ref{tab:3}. The solid line shows the mean value of $H(z)$. It is clear that using joint datasets gives raise to better fit to the data.}

\end{minipage}
\hspace{0.5cm}
\begin{minipage}[b]{0.5\linewidth}
\centering
\includegraphics[width=8cm,height=6cm,angle=0]{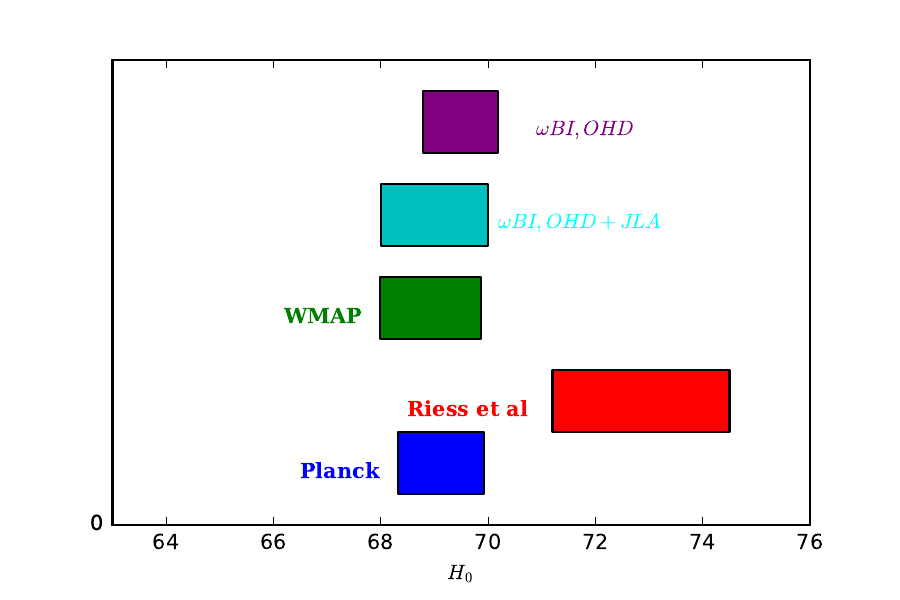}
\caption{\label{fig3} Schematic representation of $H_{0}$ (at $68\%$ CL) for flat $\omega$BI model ($OHD$(purple color) and $OHD+JLA$(cyan color)). Constraints from the direct measurement by Riess et al. (2016) (red color) WMAP (green color), and the Planck Collaboration (2015) (blue color) are also shown.}

\end{minipage}
\end{figure}
%%%%%%%%%%%%%%%%%%%%%%%%%%%%%%%%%%%%%%%%%%%%%%%%%%%%%%%%%%%%%%%%%%%%%%%%%%%%%%%%%%%%%%%%%%%%%%%%%%%%%%%
%%%%%%%%%%%%%%%%%%%%%%%%%%%%%%%%%%%%%%%%%  Figure4  %%%%%%%%%%%%%%%%%%%%%%%%%%%%%%%%%%%%%%%%%%%%%%%%%%%
\begin{figure}[h!]
\centering
\begin{subfigure}[b]{0.33\textwidth}
\centering
\includegraphics[width=\textwidth]{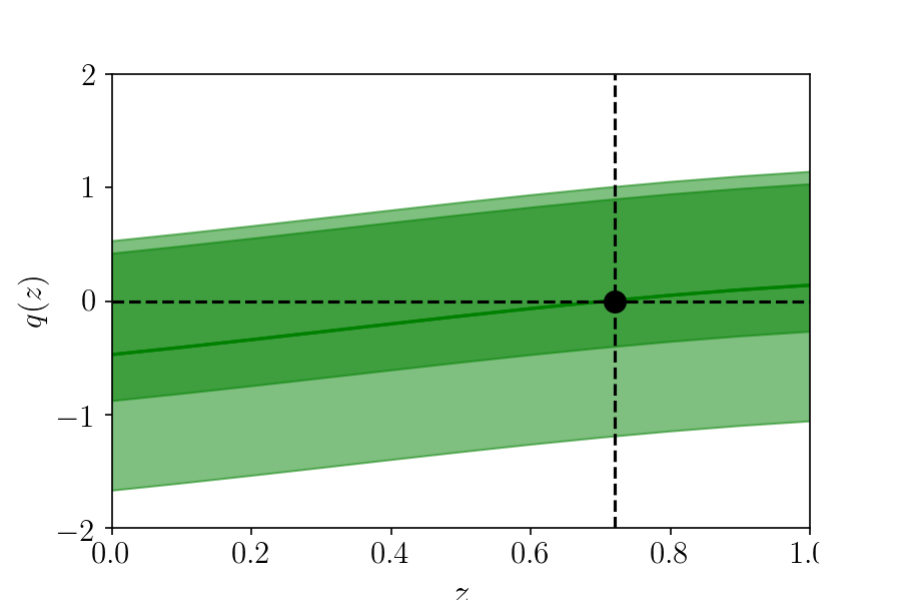}
\caption{\label{fig4a} OHD}
		
\end{subfigure}%
\begin{subfigure}[b]{0.33\textwidth}
\centering
\includegraphics[width=\textwidth]{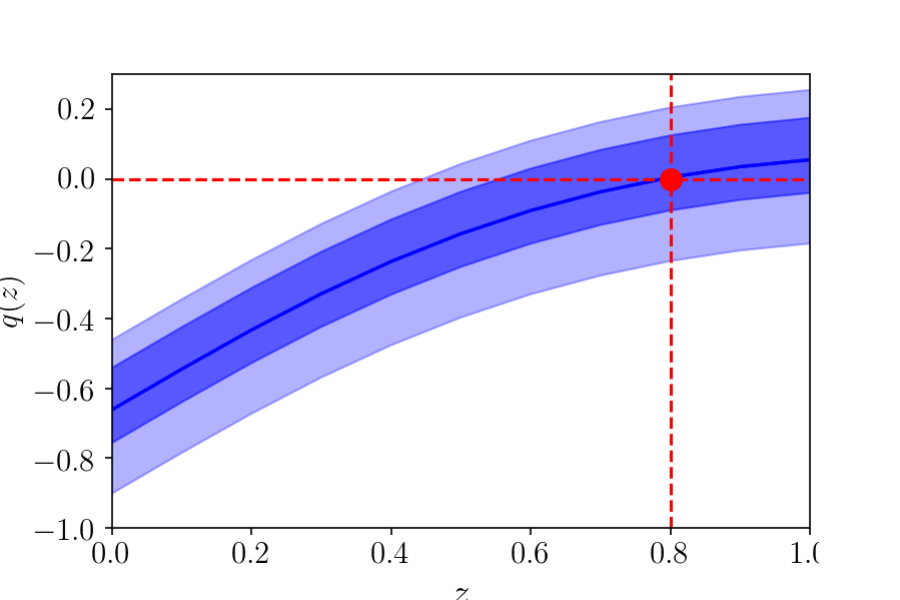}
\caption{\label{fig4b} JLA}
		
\end{subfigure}
\begin{subfigure}[b]{0.33\textwidth}
\centering
\includegraphics[width=\textwidth]{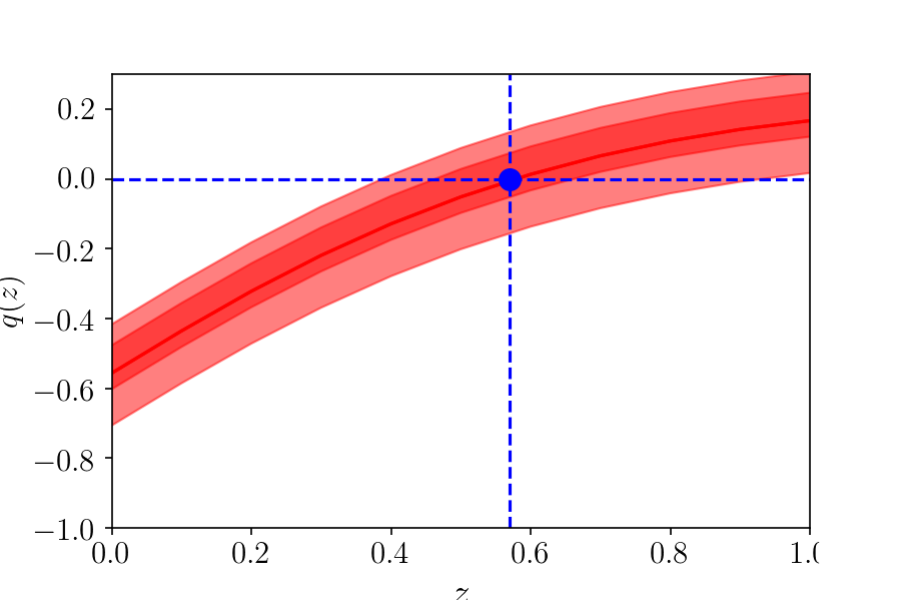}
\caption{\label{fig4c} OHD+JLA}
		
\end{subfigure}
\caption{\label{fig4} Plots of the deceleration parameter of $\omega$BI model using: (a) Hubble (OHD), (b) SNIa (JLA), and (c) OHD + JLA combination data. The solid lines show the mean value of $q(z)$ and filled circles indicate the best fit value of deceleration parameter at $z_{t}$ in each figure.}
	
\end{figure}
\begin{figure}[h!]
\centering
\begin{subfigure}[b]{0.33\textwidth}
\centering
\includegraphics[width=\textwidth]{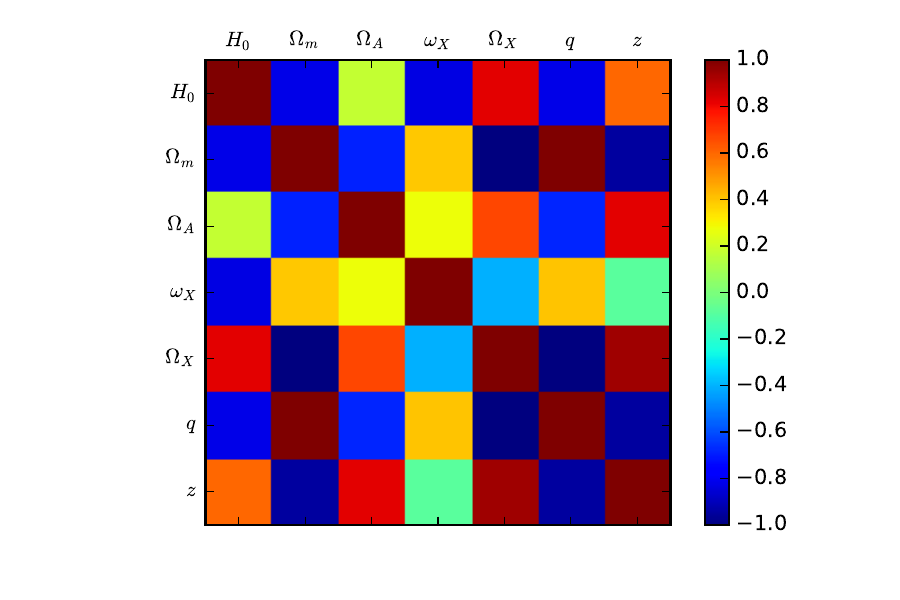}
\caption{\label{fig5a} OHD}

\end{subfigure}%
\begin{subfigure}[b]{0.33\textwidth}
\centering
\includegraphics[width=\textwidth]{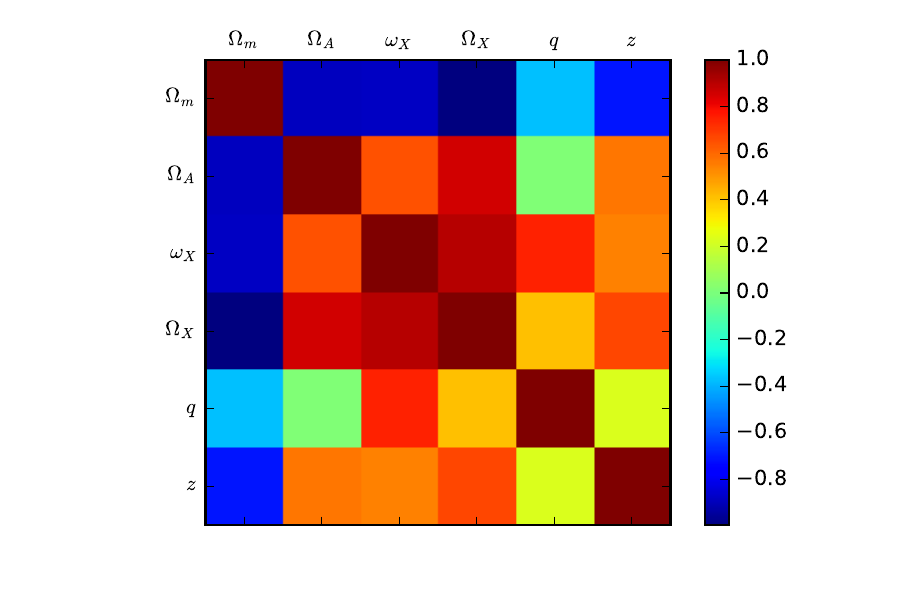}
\caption{\label{fig5b} JLA}

\end{subfigure}
\begin{subfigure}[b]{0.33\textwidth}
\centering
\includegraphics[width=\textwidth]{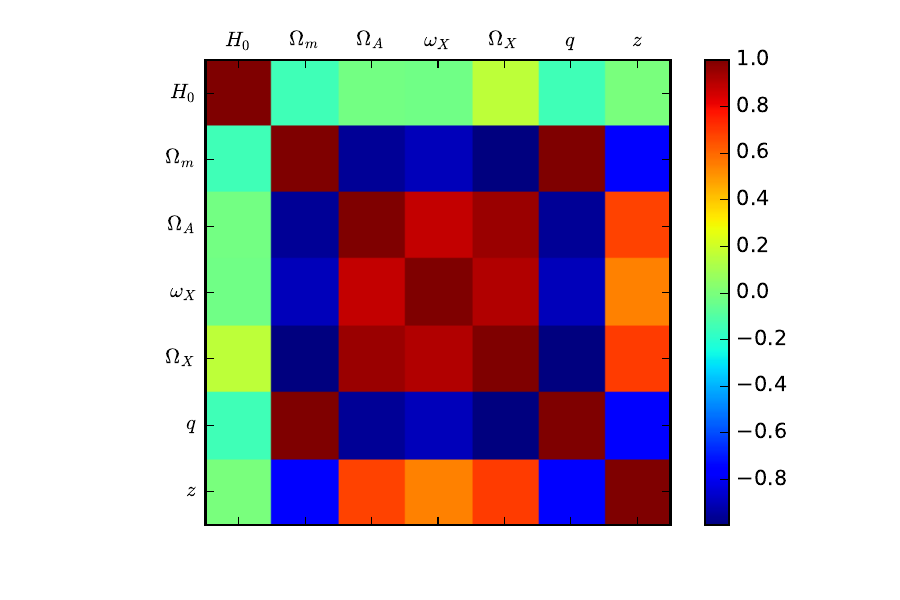}
\caption{\label{fig5c} OHD+JLA}

\end{subfigure}
\caption{\label{fig5} Plots of correlation matrix of $\omega$BI model using: (a) Hubble (OHD), (b) SNIa (JLA), and (c) OHD + JLA combination datasets. The color bars share the same scale.}

\end{figure}

%%%%%%%%%%%%%%%%%%%%%%%%%%%%%%%%%%%%%%%%%%%%%%%%%%%%%%%%%%%%%%%%%%%%%%%%%%%%%%%%%%%%%%%%%%%%%%%%%%%%%%%%%%%%%
%%%%%%%%%%%%%%%%%%%%%%%%%%%%%%%%%%%%%%%%%%%%%%%%%%%%%%%%%%%%%%%%%%%%%%%%%%%%%%%%%%%%%%%%%%%%%%%%%%%%%%%%%
A useful tool to study the degeneracy direction between estimated parameters is covariance matrix. The covariance matrix $C$ of the parameter space $\{\theta\}$ could be defined as 
\begin{equation}
\label{eq22} C_{ij}=\rho_{ij}\sigma({\theta_{i}})\sigma({\theta_{j}}),
\end{equation}
where $\rho_{ij}$ is called as the correlation coefficient between parameters $\theta_{i}$ and $\theta_{j}$. $\sigma({\theta_{i}})$ and $\sigma({\theta_{j}})$ are the $1\sigma$ uncertainties in parameters $\theta_{i}$ and $\theta_{j}$. Note that $\rho$ varies from 0 (independent) to 1 (completely correlated). We can estimate the covariance matrix $C$ from the MCMCs. Figure.~\ref{fig5} depicts the correlation matrix for OHD (fig.~\ref{fig5a}), JLA (fig.~\ref{fig5b}), and OHD+JLA (fig.~\ref{fig5c}) . It is clear that when we apply joint combination, the correlations between estimated parameters increase.
%%%%%%%%%%%%%%%%%%%%%%%%%%%%%%%%%%%%%%%%%%%%%%%%%%%%%%%%%%%%%%%%%%%%%%%%%%%%%%%%%%%%%%%%%%%%%%%%%%%%%%%
%%%%%%%%%%%%%%%%%%%%%%%%%%%%%%%%%%%%%%%%%  Figure5  %%%%%%%%%%%%%%%%%%%%%%%%%%%%%%%%%%%%%%%%%%%%%%%%%%%
%%%%%%%%%%%%%%%%%%%%%%%%%%%%%%%%%%%%%%%%%%%%%%%%%%%%%%%%%%%%%%%%%%%%%%%%%%%%%%%%%%%%%%%%%%
\section{Goodness of Fit}
\label{sec: God}
In this section we used the Akaike information criterion (AIC) \cite{ref56}, defined as $\mbox{AIC}=-2\mbox{log}\mathcal{L}_{max}+2N$ and Bayes factor \cite{ref57} defined as, $\Psi=\mathcal{L}_{max}^{M_{1}}/\mathcal{L}_{max}^{M_{2}}$ to better compare the considered model (note that $\mathcal{L}_{max}$ is the maximum likelihood and $N$ indicates the number of parameters). In AIC method the preference is given to the model with the lowest AIC. On this bases,$0 \leqslant(\bigtriangleup AIC) \leqslant 2$ indicates strong evidence for the model whereas $(\bigtriangleup AIC)\leq 7$ gives moderate support for the model.
%%%%%%%%%%%%%%%%%%%%%%%%%%%%%%%%%%%%%%%%%%%%Table 1%%%%%%%%%%%%%%%%%%%%%%%%%%%%%%%%%%%%%%%%%%%%%%%%%%%%%%%%%%%
%%%%%%%%%%%%%%%%%%%%%%%%%%%%%%%%%%%%%%%%%%%%%%%%%%%%%%%%%%%%%%%%%%%%%%%%%%%%%%%%%%%%%%%%%%%%%%%%%%%%%%
\begin{table}[ht]
\caption{Comparison of the cosmological models by $\bigtriangleup\mbox{AIC}=\mbox{AIC}_{\omega\mbox{BI}}-\mbox{AIC}_{\omega\mbox{CDM}}$, $\Psi=\mathcal{L}_{max}^{\omega\mbox{CDM}}/\mathcal{L}_{max}^{\omega\mbox{BI}}$ and Bayesian $p$-values, using individual datasets and their joint combination. Here it should be noted that the number of the free parameters for $\omega \mbox{BI}$ and $\omega\mbox{CDM}$ models are 4 \& 3, respectively. The degrees of freedom for each model is given by $N=n-m$, where $n$ is the number of data-points in each dataset and $m$ is the number of the free parameters in each model.}
\centering
\begin{tabular} {|>{\columncolor[gray]{0.8}}cccc|}
\hline
Parameter    &OHD & SNIa(JLA) & OHD+SNIa  \\[0.5ex]
\hline

$\chi^{2}_{\omega\mbox{CDM}}$ &  $17.95$ &   $4605.16$ &  $4621.74$\\[0.3cm] 
		
$\chi^{2}_{\omega \mbox{BI}}$ &  $16.46$   & $4604.10$ &$4621.75$\\[0.3cm]
		
$\bigtriangleup\mbox{AIC}$ &  $0.51$   & $0.91$ & $2.012$\\[0.3cm]
		
$\Psi$ & $0.475$ & $0.588$ & $1.001$ \\[0.3cm]
		
$\mbox{p}_{\omega\mbox{CDM}}$ & $0.52$ & $0.67$ & $0.80$ \\[0.3cm]

$\mbox{p}_{\omega \mbox{BI}}$ & $0.63$ & $0.66$ & $0.83$ \\[0.3cm]

$\mbox{N}_{\omega\mbox{CDM}}$ & $33$ & $737$ & $770$ \\[0.3cm]

$\mbox{N}_{\omega\mbox{BI}}$ & $32$ & $736$ & $771$ \\[0.5ex]
\hline
\end{tabular}
\label{tab:5}
\end{table}
%%%%%%%%%%%%%%%%%%%%%%%%%%%%%%%%%%%%%%%%%%%%%%%%%%%%%%%%%%%%%%%%%%%%%%%%%%%%%%%%%%%%%%%%%%%%%%%%%%%%%%%%
 In other hand, Bayes factor represents the odds for the model $M_{1}$ against the alternative model $M_{2}$. According to Jeffreys \cite{ref57}, whilst odds lower than $1:10$ refer to a strong evidence against $M_{1}$, in reversed, odds bigger than $10 : 1$ indicate a strong evidence against $M_{2}$ (we compare our model with flat $\omega\mbox{CDM}$ model). Furthermore, we calculate $p$-values for both OHD and JLA datasetes as well as their combination to indicate how incompatible the data are with our model. In this regard, using the above mentioned parameters, we have compared our model with $\omega\mbox{CDM}$ model in Table~\ref{tab:5}. From this table, it is clear that the results of $\bigtriangleup\mbox{AIC}$ and $\Psi$ indicate strong evidence for the model $\omega \mbox{BI}$ with respect to $\omega\mbox{CDM}$ model (note that $\omega \mbox{BI}$ model has one parameter more than flat $\omega\mbox{CDM}$ model). From the value of $\chi^{2}$ we observe that the $\omega \mbox{BI}$ model is performing as well as the flat $\omega\mbox{CDM}$ model in fitting the OHD and SNIa data. In other hand, we have presented the calculated p-value for both models in Table~\ref{tab:5}. As mentioned by Gelman et al \cite{ref58} the acceptable range for $p$-value of a model to be compatible with data is $0.05\leqslant p\leqslant0.95$, therefore, our obtained results for Bayesian $p$-value show that non of the $\omega \mbox{BI}$ and flat $\omega\mbox{CDM}$ models rule out by OHD, JLA and their joint datasets.
%%%%%%%%%%%%%%%%%%%%%%%%%%%%%%%%%%%%%%%%%%%%%%%%%%%%%%%%%%%%%%%%%%%%%%%%%%%%%%%%%%%%%%%%%%%%%%%%%%%%%%%%%%%%%
%%%%%%%%%%%%%%%%%%%%%%%%%%%%%%%%%%%%%%%%%%%%%%%%%%%%%%%%%%%%%%%%%%%%%%%%%%%%%%%%%%%%%%%%%%%%%%%%%%%%%%%%%
\section{Concluding Remarks}
\label{sec: con}
According to the recent observations, there is a tiny difference between intensities of microwaves coming from different directions of the sky. This fact motivates us to study the universe in the scope of anisotropic Bianchi type I space-time in such a way to describe our universe in more realistic situation with respect to FRW universe. We considered two independent observational datasetes namely OHD and JLA as well as their joint combination to constraint $\omega$BI dark energy model which is inherently a flat space-time. We compare our results with recent results reported by 9 years WMAP \cite{ref10} and Planck (2015) collaboration \cite{ref11}. It is found that for OHD or JLA datasets alone, at $68\%$ CL, the estimated value of the dark energy EOS parameter varies from quintessences to phantom region whereas for the joint OHD+JLA dataset the estimated EOS parameter only varies in phantom region. From Table~\ref{tab:2}, it is observed that both 9 years WMAP \& Planck (2015) predict the possibility of DE to lie in the phantom region. Using Planck 2015 TT + lowP + lensing + SNIa + BAO + $H(z)$ + $f_{\sigma_{8}}$ data, Parke \& Ratra \cite{ref59} have recently obtained $\omega_{X}= -0.994 \pm 0.033$ and $\Omega_{m}=0.3034 \pm 0.0073$ for flat XCDM model. Generally, comparing these results with those reported in Table~\ref{tab:4}, one can conclude that using JLA dataset gives more tighten constraints on $\omega$BI model with respect to OHD dataset. From the joint analysis, it is obtained that the anisotropy parameter of $\omega$BI model vary in the range $-38\times10^{-4}\leq\Omega_{A}\leq-16\times10^{-4}$ at $68\%$ confident level. This result is not in good agreement with what obtained from recent CMB observations which indicating that the anisotropy parameter is of order $\sim 10^{-5}$. In fact, this parameter is important in the study of the early universe i.e at high redshifts when the anisotropy plays more effective role in the structure formation of our universe.\\

Our statistical analysis shows that the estimated value of $H_{0}$ for joint OHD+JLA dataset is in excellent agreement with recent observations from WMAP and Planck 2015 collaboration but it has meaningful deviation from Riess et al\cite{ref51}. It is worth to mention that the measurement of the Hubble parameter at high redshifts will be possible by detecting the Sandage-Loeb signal (SL signal)\cite{ref60,ref61}. For example, the undergoing project CODEX (COsmic Dynamics and EXo-earth experiment)\footnote{http://www.iac.es/proyecto/codex/} aims to detect the SL signal with the European Extremely Large Telescope (E-ELT)\footnote{http://www.eso.org/public/teles-instr/e-elt/}.\\The age ($t_{0}$), transition redshift ($z_{t}$), and deceleration parameter ($q$) are seen to be estimated much better to the joint HOD+JLA with respect to any individual dataset. These parameters are in good agreement with those obtained in Refs \cite{ref34,ref38,ref62}. It is worth noting that other data, such as BAO, growth factor, or CMB anisotropy data can tighten the constraints on the parameters of $\omega$BI model. Therefore, constraining this model by using other datasets specially CMB anisotropy in conjunction with $H(z)$ and SNIa datasets would be of interest.
%%%%%%%%%%%%%%%%%%%%%%%%%%%%%%%%%%%%%%%%%%%%%%%%%%%%%%%%%%%%%%%%%%%%%%%%%%%%%%%%%%%%%%%%%%%%%%%%%%%%%%%%%%
%%%%%%%%%%%%%%%%%%%%%%%%%%%%%%%%%%%%%%%%%%%%%%%%%%%%%%%%%%%%%%%%%%%%%%%%%%%%%%%%%%%%%%%%%%%%%%%%%%%%%%%%%%

\section*{ACKNOWLEDGMENT}
Authors are grateful to the anonymous referee for useful comments and suggestions. We are also grateful to Professor Bharat Ratra for critical review of the manuscript prior to submission. Also, we are thankful to Anil Kumar Yadav for fruitful discussions.
%%%%%%%%%%%%%%%%%%%%%%%%%%%%%%%%%%%%%%%%%%%%%%%%%%%%%%%%%%%%%%%%%%%%%%%%%%%%%%%%%%%%%%%%%%%

\end{document}